\documentclass[aps,pre,showpacs,amsmath,amssymb,amsfonts,superscriptaddress,lengthcheck,twocolumn,longbibliography]{revtex4-1}
\usepackage{graphicx}
\usepackage{subfigure}
\usepackage{amsthm}
\usepackage{verbatim}
\usepackage{dcolumn}
\usepackage{bm}
\usepackage{epsf}
\usepackage{color}
\usepackage[colorlinks=true,citecolor=blue,linkcolor=blue,urlcolor=blue]{hyperref}%
\usepackage{xcolor}
\usepackage{dsfont}

\newcommand{\e}[1]{\exp{\left(#1\right)}}
\newcommand{\lo}[1]{\ln{\left(#1\right)}}

\newcommand{\bla}{bla\\bla\\bla\\bla\\bla}

\newcommand{\mrm}[1]{\mathrm{#1}}

\begin{document}

\title{Efficiency of harmonic quantum Otto engines at maximal power}

\author{Sebastian Deffner}
\email{deffner@umbc.edu}
\affiliation{Department of Physics, University of Maryland Baltimore County, Baltimore, MD 21250, USA}

\date{\today}

\begin{abstract}
Recent experimental breakthroughs produced the first nano heat engines that have the potential to harness quantum resources. An instrumental question is how their performance measures up against the efficiency of classical engines. For single ion engines undergoing quantum Otto cycles it has been found that the efficiency at maximal power is given by the Curzon-Ahlborn efficiency. This is rather remarkable as the Curzon-Alhbron efficiency was originally derived for endoreversible Carnot cycles. Here, we analyze two examples of endoreversible Otto engines within the same conceptual framework as Curzon and Ahlborn's original treatment. We find that for  endoreversible Otto cycles in classical harmonic oscillators the efficiency at maximal power is, indeed, given by the Curzon-Ahlborn efficiency. However, we also find that the efficiency of Otto engines made of quantum harmonic oscillators is significantly larger.
\end{abstract}

\maketitle

\section{Introdcution}

It is a standard exercise of thermodynamics to compute the efficiency of engines, i.e., to determine the relative work output for devices undergoing cyclic transformations on the thermodynamic manifold \cite{Callen1985}. Like few other applications the study of heat engines illustrates the versatility of thermodynamic concepts, since universally valid bounds can be obtained purely from macroscopic, phenomenological knowledge about physical systems. However, all ideal cycles, such as the Carnot, Stirling, Otto, Diesel, etc. cycles are only of limited practical importance, as they are comprised of quasistatic, infinitely slow state transformations. Therefore, the power output of an ideal engine is strictly zero \cite{Callen1985}.

All real engines operate in finite time, and thus their working medium is almost never in equilibrium with the environment. Moreover, a more practical question is to determine the efficiency at maximal power output, rather than focusing only at the ideal, maximal efficiency (at zero power). In a seminal paper \cite{Curzon1975}, Curzon and Ahlborn tackled this problem within the framework of \emph{endoreversible thermodynamics} \cite{Hoffmann1997}.

At the core of endoreversible thermodynamics is the idea of \emph{local equilibrium}: Imagine an engine, whose working medium is in a state of thermal equilibrium of temperature $T$. However, $T$ is not equal to the temperature of the environment, $T_\mrm{bath}$, and thus there is a temperature gradient at the boundaries of the engine. One then studies the engine as it slowly undergoes a cyclic state transformation, where slow means that the working medium remains \emph{locally} in equilibrium at all times. However, since the cycle does operate in finite time, the working medium never fully equilibrates with the environment. Therefore, from the point of view of the environment the device undergoes an irreversible cycle. Such state transformations are called \emph{endoreversible} \cite{Hoffmann1997}, which means that locally the transformation is reversible, but globally irreversible. 

Curzon and Ahlborn showed \cite{Curzon1975} that the efficiency of a Carnot engine undergoing an endoreversible cycle at maximal power is given by,
\begin{equation}
\label{eq:CA}
\eta_\mrm{CA}=1-\sqrt{\frac{T_c}{T_h}}\,,
\end{equation}
where $T_c$ and $T_h$ are the temperatures of the cold and hot reservoirs, respectively. Remarkably, it has been found that $\eta_\mrm{CA}$ \eqref{eq:CA} is also assumed by many, physically different engines at maximal power, such as an endoreversible Otto engine with an ideal gas as working medium \cite{Leff1987}, the endoreversible Stirling cycle \cite{Erbay1997}, Otto engines in open quantum systems in the quasistatic limit \cite{Rezek2006}, or a single ion in a harmonic trap undergoing a quantum Otto cycle \cite{Abah2012,Rossnagel2014}. On the other hand, it also has been shown that whether or not a finite time Carnot cycle assumes $\eta_\mrm{CA}$ is determined by the ``symmetry'' of dissipation \cite{Esposito2010} and the efficiency of an Otto engine working with a single Brownian particle in a harmonic trap is determined by the specific parameterization of the trap's stiffness \cite{Bonanca2018}.

In particular, the recent experimental breakthroughs in the implementation of nanosized heat engines \cite{Rossnagel2016,Klaers2017} that could principally exploit quantum resources \cite{Scovil1959,Scully2002,Scully2003,Scully2011,Zhang2014PRL,Gardas2015,Hardal2015,Cavina2017,Roulet2018,Cherubim2018,Niedenzu2018,Ronzani2018} pose the question whether their behavior can be universally characterized. For instance, Ref.~\cite{Rezek2006} suggested that to describe the efficiency at maximal power $\eta_\mrm{CA}$ could be such a universal result, at least for a class of engines. However, the Curzon-Ahlborn efficiency \eqref{eq:CA} was originally derived for endoreversible Carnot cycles, which is independent on the nature of the working medium. On the other hand, a standard textbook exercise shows that the \emph{Otto efficiency} is dependent on the equation of state, i.e., on the specific working medium \cite{Callen1985}. Therefore, it would actually be more natural to expect that the efficiency at maximal power strongly depends on the nature of working medium. Similar conclusions have been drawn, for instance, in the thermodynamic analysis of photovolatic cells \cite{Scully2010,Dorfman2013,Einax2014}.

In addition, the \emph{quantum Otto cycle} is typically comprised of two thermalization and two unitary strokes \cite{Quan2007,Kosloff2013,Kosloff2017}. For cycles involving only unitary strokes \cite{Abah2012,Rossnagel2014} the assumption of local equilibrium is almost never justified, and thus it becomes even more remarkable that at maximal power output a quantum Otto cycle in a parametric, harmonic oscillator operates with the Curzon-Ahlborn efficiency \cite{Abah2012,Rossnagel2014}. Also see Ref.~\cite{Rezek2006} for a more detailed treatment from open quantum dynamics. Therefore, the question arises whether this is a peculiarity of the quantum Otto cycle in the harmonic oscillator, or whether there is something more fundamental and universal about $\eta_\mrm{CA}$.

The purpose of the present work is to revisit these longstanding questions and study the endoreversible Otto cycle in a conceptually simple and pedagogical approach similar to Curzon and Ahlborn's original treatment \cite{Curzon1975}. To this end, we compute the efficiency at maximal power for two examples of endoreversible Otto engines. We start with a classical version, for which the working medium is a single Brownian particle in a harmonic trap. Maximizing the power output with respect to the compression ratio, we find \emph{analytically} that the efficiency is indeed given by $\eta_\mrm{CA}$ \eqref{eq:CA}. As a second example we study a quantum engine, whose working medium is a quantum harmonic oscillator ultraweakly coupled to the thermal environment. We find that in this case the efficiency is larger than  $\eta_\mrm{CA}$ \eqref{eq:CA}, which demonstrates that the Curzon-Ahlborn efficiency is \emph{not} universal at maximal power. An advantage of the present treatment is that it is somewhat more pedagogical than earlier works on the topic. The present derivation is entirely based on the phenomenological framework of endoreversible thermodynamics. Thus, e.g., neither the full quantum dynamics \cite{Rezek2006} nor the linear response problem \cite{Bonanca2018} have to be solved.

\section{Carnot engine at maximal power}

We begin by briefly reviewing the main gist of Ref.~\cite{Curzon1975} and by establishing notions and notation. At that we focus on the limits and assumptions that lead to the Curzon-Ahlborn efficiency \eqref{eq:CA} for endoreversible Carnot engines. 

The ideal Carnot cycle consists of two isothermal processes during which the systems absorbs/exhausts heat and two thermodynamically adiabatic, i.e., isentropic strokes \cite{Callen1985}. During the isentropic strokes the working medium does not exchange heat with the thermal reservoirs, and thus its state can be considered to be independent of the environment. Therefore, we only have to modify the treatment of the isothermal strokes during which the working medium will be in a local equilibrium state at different temperature than the temperature of the hot and cold reservoir, respectively.

In particular, during the hot isotherm the working medium is assumed to be a little cooler than the hot environment at $T_h$. Thus, during the whole stroke the system absorbs the heat
\begin{equation}
\label{eq:heat_hot}
Q_h=\lambda_h \tau_h \left(T_h-T_{hw}\right)\,,
\end{equation}
where $\tau_h$ is the stroke time, $T_{h,w}$ is the temperature of the working medium, and $\lambda_h$ is a constant depending on thickness and thermal conductivity of the boundary separating working medium and environment. Note that Eq.~\eqref{eq:heat_hot} is nothing else but a discretized version of Fourier's law for heat conduction \cite{Callen1985}. We will see shortly that for Otto cycles the rate of heat flux can no longer be assumed to be constant, since we need to account for the change in temperature during the isochoric strokes.

Similarly, during the cold isotherm the system is a little warmer than the cold reservoir at $T_c$. Hence, the exhausted heat can be written as
\begin{equation}
\label{eq:heat_cold}
Q_c=\lambda_c \tau_c \left(T_{cw}-T_{c}\right)
\end{equation}
where $\lambda_c$ is the cold heat transfer coefficient. 

As mentioned above, the adiabatic strokes are unmodified, but note that the cycle is taken to be reversible with respect to the \emph{local temperatures} of the working medium. Hence, we can write
\begin{equation}
\label{eq:entropy}
\Delta S_h=-\Delta S_c\quad\mrm{and\,\, thus}\quad\frac{Q_h}{T_{hw}}=\frac{Q_c}{T_{cw}}\,.
\end{equation}
Equation~\eqref{eq:entropy} allows to relate the stroke times $\tau_h$ and $\tau_c$ to the heat transfer coefficients $\lambda_h$ and $\lambda_c$.

We are now interested in determining the efficiency at maximal power. To this end, we write the power output of the cycle as
\begin{equation}
\label{eq:power}
P(\delta T_h,\delta T_c)=\frac{Q_h-Q_c}{\gamma (\tau_h+\tau_c)}
\end{equation}
where $\delta T_h=T_h-T_{hw}$ and $\delta T_c=T_{cw}-T_{c}$. In Eq.~\eqref{eq:power} we introduced the total cycle time $\tau_\mrm{cyc}=\gamma (\tau_h+\tau_c)$, and thus $\gamma\equiv \tau_\mrm{cyc}/(\tau_h+\tau_c)$. Note that this neglects any explicit dependence of the analysis on the lengths of the adiabatic strokes. We exclusively focus on the isotherms, i.e, on the temperature differences between working medium and the hot and cold reservoirs.

It is worth emphasizing that in the present problem we have four free parameters, namely hot and cold temperatures of the working substance, $T_{hw}$ and $T_{cw}$, and the stroke times $\tau_h$ and $\tau_c$. The balance equation for the entropy \eqref{eq:entropy} allows to eliminate two of these, and Curzon and Ahlborn chose to eliminate  $\tau_h$ and $\tau_c$ \cite{Curzon1975}. 

Thus, we maximize the power $P(\delta T_h,\delta T_c)$ as a function of the difference in temperatures between working substance and environment. After a few lines of algebra one obtains \cite{Curzon1975},
\begin{equation}
\label{eq:power_max}
P_\mrm{max}=\frac{\lambda_h\lambda_c}{\gamma}\left(\frac{\sqrt{T_h}-\sqrt{T_c}}{\sqrt{\lambda_h}+\sqrt{\lambda_c}}\right)^2\,,
\end{equation}
where the maximum is assumed for
\begin{equation}
\label{eq:deltaT_max}
\frac{\delta T_h}{T_h}=\frac{1-\sqrt{T_c/T_h}}{1+\sqrt{\lambda_h/\lambda_c}}\quad\mrm{and}\quad \frac{\delta T_c}{T_c}=\frac{\sqrt{T_h/T_c}-1}{1+\sqrt{\lambda_c/\lambda_h}}
\end{equation}
From these expressions we can now compute the efficiency. We have,
\begin{equation}
\label{eq:efficency}
\eta=\frac{Q_h-Q_c}{Q_h}=1-\frac{T_{cw}}{T_{hw}}=1-\frac{T_c+\delta T_c}{T_h-\delta T_h}
\end{equation}
where we used Eq.~\eqref{eq:entropy}. Thus, the efficiency of an endoreversible Carnot cycle at maximal power output becomes
\begin{equation}
\label{eq:efficency_CA}
\eta_{CA}=1-\sqrt{\frac{T_c}{T_h}}\,,
\end{equation}
which only depends on the temperatures of the hot and cold reservoirs.

In the following, we will apply exactly the same reasoning to the endoreversible Otto cycle.

\section{Endoreversible Otto cycle}

The standard Otto cycle is a four-stroke cycle consisting of isentropic compression, isochoric heating, isentropic expansion, and ischoric cooling \cite{Callen1985}. Thus, we have in the endoreversible regime:

\paragraph*{Isentropic compression}
During the isentropic strokes the working substance does not exchange heat with the environment. Therefore, the thermodynamic state of the working substance can be considered independent of the environment, and the endoreversible description is identical to the equilibrium cycle. From the first law of thermodynamics, $\Delta E=Q+W$, we have,
\begin{equation}
Q_\mrm{comp}=0\quad\mrm{and}\quad W_\mrm{comp}=E(T_2,\omega_2)-E(T_1,\omega_1)
\end{equation}
where $Q_\mrm{comp}$ is the heat exchanged, and $W_\mrm{comp}$ is the work performed during the compression. Moreover, $\omega$ denotes the work parameter, such as the inverse volume of a piston or the  frequency of a harmonic oscillator \eqref{eq:harm}.

\paragraph*{Isochoric heating}

During the isochoric strokes the work parameter is held constant, and the system exchanges heat with the environment. Thus, we have for isochoric heating 
\begin{equation}
Q_h=E(T_3,\omega_2)-E(T_2,\omega_2)\quad\mrm{and}\quad W_h=0\,.
\end{equation}

In complete analogy to Curzon and Ahlborn's original analysis \cite{Curzon1975} we now assume that the working substance is in a state of local equilibrium, but also that the working substance never fully equilibrates with the hot reservoir. Therefore, we can write
\begin{equation}
T(0)=T_2\quad\mrm{and}\quad T(\tau_h)=T_3\quad\mrm{with}\quad T_2<T_3 \leq T_h\,,
\end{equation}
where as before $\tau_h$ is the duration of the stroke.

Note that in contrast to the Carnot cycle the Otto cycle does not involve isothermal strokes, and, hence, the rate of heat flux is not constant. Rather, we have to explicitly account for the change in temperature from $T_2$ to $T_3$. To this end, Eq.~\eqref{eq:heat_hot} is replaced by Fourier's law \cite{Callen1985},
\begin{equation}
\label{eq:fourier_hot}
\frac{d T}{dt}=-\alpha_h \left(T(t)-T_h\right)
\end{equation}
where $\alpha_h$ is a constant depending on the heat conductivity and heat capacity of the working substance. 

Equation~\eqref{eq:fourier_hot} can be solved exactly, and we obtain the relation
\begin{equation}
\label{eq:rel_hot}
T_3-T_h=\left(T_2-T_h\right)\,\e{-\alpha_h \tau_h}\,.
\end{equation}
In the following, we will see that Eq.~\eqref{eq:rel_hot} is instrumental in reducing the number of free parameters.

\paragraph*{Isentropic expansion}
 In complete analogy to the compression, we have for the isentropic  expansion,
\begin{equation}
Q_\mrm{exp}=0\quad\mrm{and}\quad W_\mrm{exp}=E(T_4,\omega_1)-E(T_3,\omega_2)\,.
\end{equation}

\paragraph*{Isochoric cooling} 

Heat and work during the isochoric cooling read,
\begin{equation}
Q_c=E(T_1,\omega_1)-E(T_4,\omega_1)\quad\mrm{and}\quad W_c=0\,,
\end{equation}
where we now have
\begin{equation}
T(0)=T_4\quad\mrm{and}\quad T(\tau_c)=T_1\quad\mrm{with}\quad T_4>T_1 \geq T_c\,.
\end{equation}

Similarly to above \eqref{eq:fourier_hot} the heat transfer is described by Fourier's law
\begin{equation}
\label{eq:fourier_cold} 
\frac{d T}{dt}=-\alpha_c \left(T(t)-T_c\right)\,,
\end{equation}
where $\alpha_c$ is a constant characteristic for the cold stroke. From the solution of Eq.~\eqref{eq:fourier_cold} we now obtain
\begin{equation}
\label{eq:rel_cold}
T_1-T_c=\left(T_4-T_c\right)\,\e{-\alpha_c \tau_c}\,,
\end{equation}
which properly describes the decrease in temperature from $T_4$ back to $T_1$.

\section{Classical harmonic engine}

To continue the analysis we now need to specify the internal energy $E$. As a first example, we consider a classical Brownian particle trapped in a harmonic oscillator. The bare Hamiltonian reads,
\begin{equation}
\label{eq:harm}
H(p,x)=\frac{p^2}{2m}+\frac{1}{2}m \omega^2 x^2\,,
\end{equation}
where $m$ is the mass of the particle. 

For a particle in thermal equilibrium the Gibbs entropy, $S$, and the internal energy, $E$, are
\begin{equation}
\label{eq:harm_ent}
\frac{S}{k_B}=1+\lo{\frac{k_B T}{\hbar\omega}}\quad\mrm{and}\quad E=k_B T\,,
\end{equation}
where we introduced Boltzmann's constant, $k_B$. 

Note, that from Eq.~\eqref{eq:harm_ent} we obtain a relation between the frequencies, $\omega_1$ and $\omega_2$ and the four temperatures, $T_1$, $T_2$, $T_3$, and $T_4$. To this end, consider the isentropic strokes, for which we have
\begin{equation}
S(T_2,\omega_2)=S(T_1,\omega_1)\quad\mrm{and}\quad S(T_4,\omega_1)=S(T_3,\omega_2)\,,
\end{equation}
which is fulfilled by
\begin{equation}
\label{eq:temp}
T_1\, \omega_2=T_2\,\omega_1\quad\mrm{and}\quad T_3\, \omega_1=T_4\,\omega_2\,.
\end{equation}

We are now equipped with all the ingredients necessary to compute the endoreversible efficiency, 
\begin{equation}
\label{eq:eff}
\eta=-\frac{W_\mrm{tot}}{Q_h}\,.
\end{equation}
In complete analogy to fully reversible cycles \cite{Callen1985}, Eq.~\eqref{eq:eff} can be written as
\begin{equation}
\label{eq:eff_temp}
\eta=1-\frac{T_4-T_1}{T_3-T_2}\,,
\end{equation}
where we used the explicit from of the internal energy $E$ \eqref{eq:harm_ent}. Further, using Eqs.~\eqref{eq:temp} the endoreversible Otto efficiency becomes
\begin{equation}
\eta=1-\frac{\omega_1}{\omega_2}\equiv1-\kappa\,,
\end{equation}
which defines the compression ratio, $\kappa$. Observe that the endoreversible efficiency takes the same form as its reversible counter part \cite{Callen1985}. However, in Eq.~\eqref{eq:eff_temp} the temperatures correspond the local equilibrium state of the working substance, and not to a \emph{global} equilibrium with the environment. 

Similarly to Curzon and Ahlborn's treatment of the endoreversible Carnot cycle \cite{Curzon1975} we now compute the efficiency for a value of $\kappa$, at which the power \eqref{eq:power} is maximal. We begin by re-writing the total work with the help of the compression ratio $\kappa$ and Eqs.~\eqref{eq:temp} as,
\begin{equation}
W_\mrm{tot}=W_\mrm{comp}+W_\mrm{exp}=\left(\kappa -1\right)\,k_B\,\left(T_2-T_3\right)\,.
\end{equation}
Further using Eq.~\eqref{eq:rel_hot} we obtain
\begin{equation}
W_\mrm{tot}=\left(\kappa -1\right)\left(1-\e{-\alpha_h\tau_h}\right)\,k_B\,\left(T_2-T_h\right)\,,
\end{equation}
which only depends on the free parameters $T_2$, $\kappa$, and $\tau_h$. Of these three, we can eliminate one more, by combing Eqs.~\eqref{eq:rel_hot} and \eqref{eq:rel_cold}, and we have
\begin{equation}
T_2=\frac{T_c\left(e^{\alpha_c\tau_c}-1\right)+\kappa\,T_h\left(1-e^{-\alpha_h\tau_h}\right)}{\kappa\left(e^{\alpha_c\tau_c}-e^{-\alpha_h\tau_h}\right)}\,.
\end{equation}
Finally, the power output \eqref{eq:power} takes the form, 
\begin{equation}
\label{eq:power_clas}
P=\frac{2(\kappa-1)\,k_B\,(T_c-\kappa\,T_h)}{\gamma\kappa (\tau_c+\tau_h)}\,\frac{\sinh{\left(\alpha_c\tau_c/2\right)}\sinh{\left(\alpha_h\tau_h/2\right)}}{\sinh{\left[(\alpha_c\tau_c+\alpha_h\tau_h)/2\right]}}\,.
\end{equation}
Remarkably the power output, $P(\kappa,\tau_h,\tau_c)$, factorizes into a contribution that only depends on the compression ratio, $\kappa$, and another term that is governed by the stroke times, $\tau_c$ and $\tau_h$,
\begin{equation}
P(\kappa,\tau_h,\tau_c)=f_1(\kappa) f_2(\tau_h,\tau_c)\,.
\end{equation}
It is then a simple exercise to show that $P(\kappa,\tau_h,\tau_c)$ is maximized for any value of $\tau_h$ and $\tau_c$ if we have,
\begin{equation}
P_\mrm{max}=P(\kappa_\mrm{max})\quad\mrm{with}\quad \kappa_\mrm{max}=\sqrt{\frac{T_c}{T_h}}\,.
\end{equation}
Therefore, the efficiency at maximal power reads,
\begin{equation}
\label{eq:eff_result}
\eta=1-\sqrt{\frac{T_c}{T_h}}\,.
\end{equation}
In conclusion, we have shown that for the classical harmonic oscillator the efficiency at maximal power of an endoreversible Otto cycle \eqref{eq:eff} is indeed given by the Curzon-Ahlborn efficiency \eqref{eq:CA}.

It is worth emphasizing that for the endoreversible Otto cycle we started with six free parameters, the four temperatures $T_1$, $T_2$, $T_3$, and $T_4$, and the two stroke times, $\tau_h$ and $\tau_c$. Of these, we succeeded in eliminating three, by explicitly using Fourier's law for the heat transfer, Eqs.~\eqref{eq:fourier_hot} and \eqref{eq:fourier_cold}, and the explicit expressions for the entropy and the internal energy \eqref{eq:harm_ent}. Therefore, one would not expect to obtain the same result \eqref{eq:eff_result} for other working substances such as the quantum harmonic oscillator.

\section{Quantum harmonic engine}

For the remainder of this analysis we will be interested in a quantum harmonic oscillator in the ultraweak coupling limit \cite{Spohn1978}. In this limit, a ``small'' quantum system interacts only weakly with a large Markovian heat bath, such that the stationary state is given by a thermal equilibrium distribution. This situation is similar to the model studied in Ref.~\cite{Rezek2006}, however in the present case we will \emph{not} have to solve the full quantum dynamics.

The equilibrium state is given by a Gibbs state, $\rho\propto \e{-H/k_B T}$, where $\rho$ is the density operator. Accordingly, the internal energy reads
\begin{equation}
\label{eq:qu_energy}
E=\frac{\hbar\omega}{2}\coth{\left(\frac{\hbar\omega }{2 k_B T}\right)}
\end{equation}
and the entropy becomes
\begin{equation}
\label{eq:qu_entropy}
\frac{S}{k_B}=\frac{\hbar\omega }{2 k_B T}\coth{\left(\frac{\hbar\omega }{2 k_B T} \right)}-\ln{\left[\frac{1}{2}\sinh{\left(\frac{\hbar\omega }{2 k_B T}\right)}\right]}\,.
\end{equation}

Despite the functional form of $S$ being more involved, we notice that the four temperatures and the two frequencies are still related by the same Eq.~\eqref{eq:temp}. Thus, it can be shown \cite{Rezek2006} that the efficiency of an endoreversible Otto cycle in a quantum harmonic oscillators also reads,
\begin{equation}
\label{eq:eff_qu}
\eta=1-\kappa\,.
\end{equation}

\begin{widetext}
Following the analogous steps that led to Eq.~\eqref{eq:power_clas} we obtain for the power output of an endoreversible quantum Otto engine,
\begin{equation}
\label{eq:power_qu}
\begin{split}
&P= \mrm{csch}{\left[\frac{\hbar\omega_2\,\kappa}{2}\,\frac {e^{\alpha_c\tau_c+\alpha_h\tau_h}-1}{T_c \left(e^{\alpha_c\tau_c}-1 \right)+\kappa T_h\,e^{\alpha_c\tau_c}\left(e^{\alpha_h\tau_h}-1\right)}\right]}\,\mrm{csch}{\left[\frac{\hbar\omega_2\,\kappa}{2}\, \frac{e^{\alpha_c\tau_c+\alpha_h\tau_h}-1}{T_c\,e^{\alpha_h\tau_h}\left(e^{\alpha_c\tau_c}-1\right)+\kappa T_h\left(e^{\alpha_h\tau_h}-1\right)}\right]}\\
&\times\frac{\hbar\omega_2}{2}\,\frac{1-\kappa}{\tau_c+\tau_h}\,\mrm{sinh}{\left[\frac{\hbar\omega_2\,\kappa}{2}\frac{\left(\kappa T_h-T_c\right)\left(e^{\alpha_c\tau_c+\alpha_h\tau_h}-1\right)\left(e^{\alpha_h\tau_h}-1\right)\left(e^{\alpha_c\tau_c}-1\right)}{\left(T_c \left(e^{\alpha_c\tau_c}-1 \right)+\kappa T_h\,e^{\alpha_c\tau_c}\left(e^{\alpha_h\tau_h}-1\right)\right)\left(T_c\,e^{\alpha_h\tau_h}\left(e^{\alpha_c\tau_c}-1\right)+\kappa T_h\left(e^{\alpha_h\tau_h}-1\right)\right)}\right]}
\end{split}
\end{equation}
where we set $k_B=1$. We immediately observe that in contrast to the classical case \eqref{eq:power_clas} the expression no longer factorizes. Consequently, the value of $\kappa$, for which $P$ is maximal does depend on the stroke times $\tau_h$ and $\tau_c$.
\end{widetext}

\begin{figure}
\centering
\includegraphics[width=.48\textwidth]{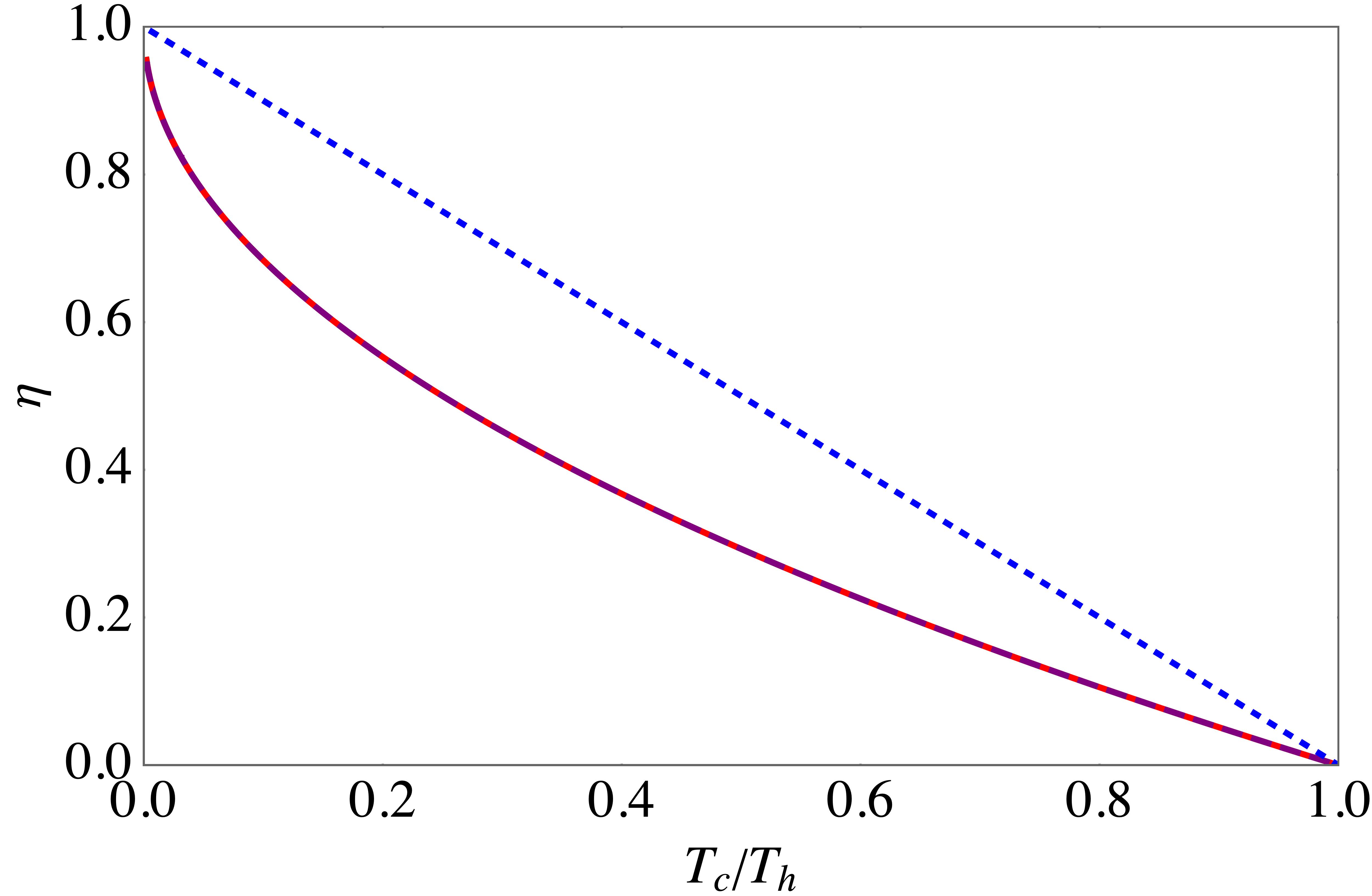}
\caption{\label{fig:efficiency_high} Efficiency of the endoreversible Otto cycle at maximal power (red, solid line), together with the Curzon-Ahlborn efficiency (purple, dashed line) and the Carnot efficiency (blue, dotted line) in the high temperature limit, $\hbar\omega_2/k_B T_c =0.1$.  Other parameters are $\alpha_c=1$, $\alpha_h=1$, and $\gamma=1$.}
\end{figure}

Due to the somewhat cumbersome expression \eqref{eq:power_qu} we chose to find the maximum of $P(\kappa,\tau_h,\tau_c)$ numerically. In Fig.~\ref{fig:efficiency_high} we illustrate our findings in the high temperature limit, $\hbar\omega_2/k_B T_c\ll 1$. Consistently with our classical example, the efficiency is given by Eq.~\eqref{eq:eff_result}, which was also found in Ref.~\cite{Rezek2006} for quasistatic cycles. It is worth emphasizing that Fig.~\ref{fig:efficiency_high} was  obtained numerically for a specific choice of parameters. However, the above, classical analysis revealed that  in the limit of high temperatures the result, namely that the efficiency at maximal power is given by the Curzon-Ahlborn efficiency \eqref{eq:eff_result},  becomes independent of all parameters but the temperatures of the hot and cold reservoirs.

\begin{figure}
\centering
\includegraphics[width=.48\textwidth]{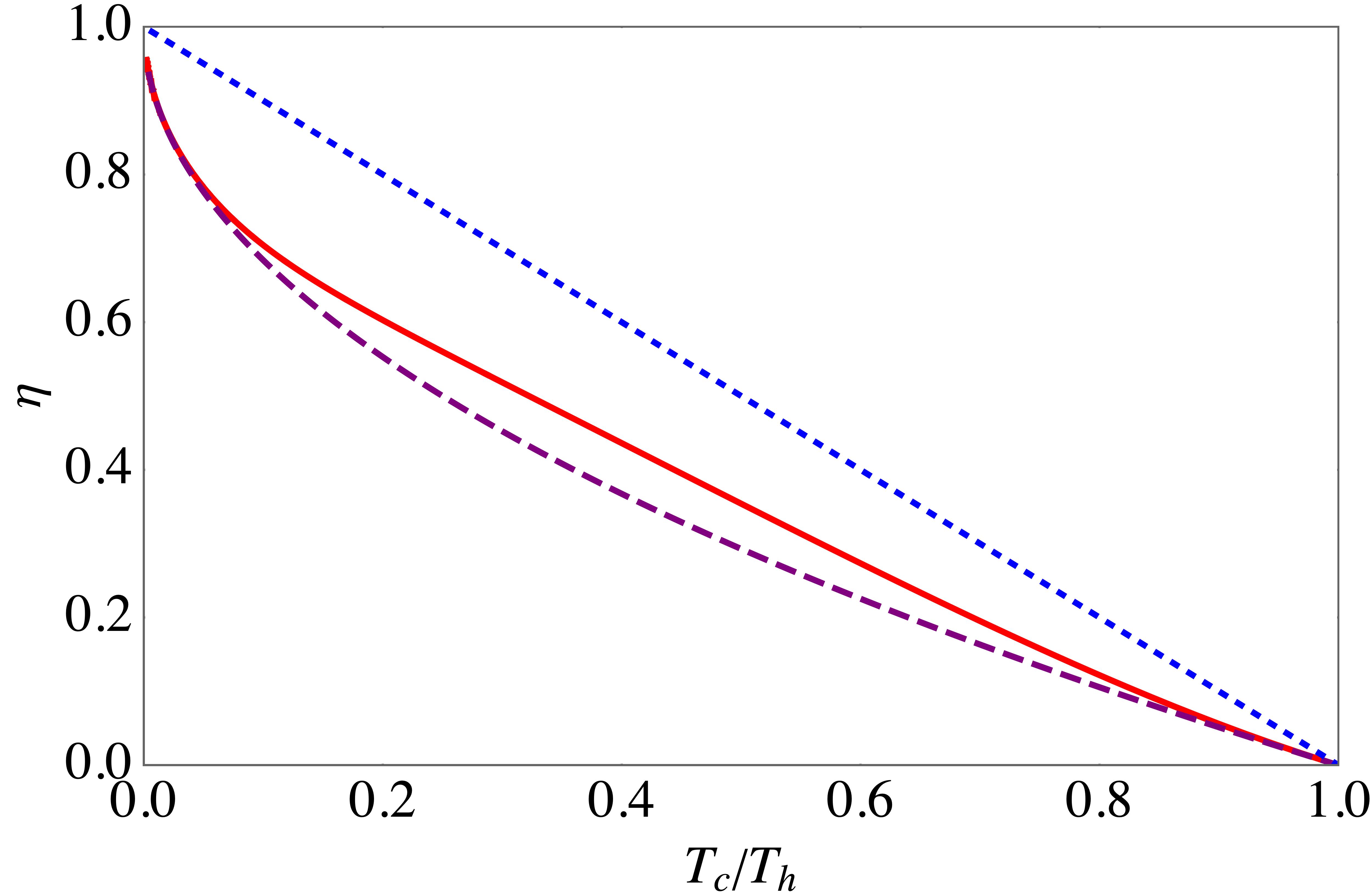}
\caption{\label{fig:efficiency_low}Efficiency of the endoreversible Otto cycle at maximal power (red, solid line), together with the Curzon-Ahlborn efficiency (purple, dashed line) and the Carnot efficiency (blue, dotted line) in the deep quantum regime, $\hbar\omega_2/k_B T_c =10$. Other parameters are $\alpha_c=1$, $\alpha_h=1$, and $\gamma=1$.}
\end{figure}

Figure~\ref{fig:efficiency_low} depicts the efficiency at maximal power \eqref{eq:eff_qu} as a function of $T_c/T_h$ in the deep quantum regime,  $\hbar\omega_2/k_B T_c\gg 1$. In this case, we find that the quantum efficiency is larger than the Curzon-Ahlborn efficiency \eqref{eq:eff_result}. From a thermodynamics' point-of-view this finding is not really surprising since already in reversible cycles the efficiency strongly depends on the equation of state.

In conclusion, we have shown explicitly that contrary to anecdotal evidence in the literature \cite{Leff1987,Rezek2006,Abah2012,Rossnagel2014,Klaers2017} the efficiency at maximal power is \emph{not} universally given by the Curzon-Ahlborn efficiency -- not even for the harmonic oscillator. The natural question now is if and how this ``quantum supremacy'' can be exploited in the design and experimental implementation of nano engines. This, however, we leave for future work.

\section{Concluding Remarks}

In the present work we have computed the efficiency at maximal power for two examples of the endoreversible Otto engine. We have found that in the case of a classical harmonic oscillator the efficiency is identical to the Curzon-Ahlborn expression originally found for endoreversible Carnot cycles. However, we have also shown that for engines operating with quantum harmonic oscillators the efficiency significantly differs from the classical expression. These findings are consistent with Refs.~\cite{Rezek2006} and \cite{Bonanca2018}, where it was argued that the efficiency should be governed by internal friction and specific driving protocols, respectively. The advantage of the present analysis is, however, that our results were obtained entirely from the phenomenological equations of endoreversible thermodynamics. Neither the quantum master equation \cite{Rezek2006} nor the linear response problem \cite{Bonanca2018} had to be solved explicitly.

Finally, we note that the present conclusions are a consequence of the deviating equations of state for the classical and quantum harmonic oscillator. More precisely, the maximal power output is governed by the different expressions for the internal energies. As such, the conclusions drawn in this work are more ``thermodynamical'' as they are ``quantum''. By this we mean, that it is entirely possible to find classical working substances, for which the efficiency at maximal power is not given by the Curzon-Ahlborn efficiency. We also have not excluded the existence of other quantum working substance, for which are described by the Curzon-Ahlborn efficiency. However, the hunt for these systems we also leave for future work.

\acknowledgements{It is a pleasure to thank Gregory Huxtable for enjoyable discussions during an early stage of this project, and Steve Campbell, Obinna Abah, and Marcus V. S. Bonan\c{c}a for many years of fruitful exchange of ideas. S.D. acknowledges support from the U.S. National Science Foundation under Grant No. CHE-1648973.}

\bibliography{quCA} 

\end{document}